\documentclass[sigconf]{acmart}

\usepackage{booktabs} 
\usepackage{float}
\usepackage{booktabs}
\usepackage{multirow}
\usepackage{soul}
\usepackage{amsmath}
\usepackage{graphicx}
\usepackage{caption}
\usepackage{subcaption}
\usepackage{adjustbox}

\DeclareMathOperator*{\argmax}{argmax} 

\setcopyright{rightsretained}

\acmDOI{10.475/123_4}

\acmConference[RecSys 2018]{ACM RecSys conference}{Oct 2018}{Vancouver, Canada}
\acmYear{2018}
\copyrightyear{2018}
\acmArticle{4}
\acmPrice{15.00}

\newcommand{\etal}{\textit{et al.}}

\begin{document}

\title{Improving Explainable Recommendations with Synthetic Reviews}

\author{Sixun Ouyang}
\affiliation{%
  \institution{Insight Centre for Data Analytics}
  \city{University College Dublin}
  \country{Ireland}}
\email{sixun.ouyang@insight-centre.org}

\author{Aonghus Lawlor}
\affiliation{
  \institution{Insight Centre for Data Analytics}
  \city{University College Dublin}
  \country{Ireland}
}
\email{aonghus.lawlor@insight-centre.org}

\author{Felipe Costa}
\affiliation{
 \institution{Aalborg University}
 \city{Aalborg}
 \country{Denmark}}
\email{fcosta@cs.aau.dk}

\author{Peter Dolog}
\affiliation{
  \institution{Aalborg University}
  \city{Aalborg}
 \country{Denmark}}
\email{dolog@cs.aau.dk}


\begin{abstract}

An important task for a recommender system to provide interpretable explanations for the user. This is important for the credibility of the system. Current interpretable recommender systems tend to focus on certain features known to be important to the user and offer their explanations in a structured form. It is well known that user generated reviews and feedback from reviewers have strong leverage over the users' decisions. On the other hand, recent text generation works have been shown to generate text of similar quality to human written text, and we aim to show that generated text can be successfully used to explain recommendations.

In this paper, we propose a framework consisting of popular review-oriented generation models aiming to create personalised explanations for recommendations. The interpretations are generated at both character and word levels. We build a dataset containing reviewers' feedback from the Amazon books review dataset. Our cross-domain experiments are designed to bridge from natural language processing to the recommender system domain. Besides language model evaluation methods, we employ DeepCoNN, a novel review-oriented recommender system using a deep neural network, to evaluate the recommendation performance of generated reviews by root mean square error (RMSE).  
We demonstrate that the synthetic personalised reviews have better recommendation performance than human written reviews. To our knowledge, this presents the first machine-generated natural language explanations for rating prediction.

\end{abstract}

%
%



\maketitle
%
\section{Introduction}

Collaborative filtering and neural network techniques are at the cornerstone of recent improvements in recommender system research. They aim to predict the actual rating that a target user gives to an item, taking into account the motivations of the user's interactions. However, predicting the ratings in this way cannot explain the users' behaviour or their motivations. Recently, reviews have been explored as additional source of information for rating prediction, due to their ability to describe the reason of associated user rating behaviour \cite{seo2017interpretable}. Moreover, Knijnenburg \etal \cite{knijnenburg2012explaining} indicate the user's reviews influence the experiences of others users, but they have not received proper attention. 

Recently, interpretable recommender systems apply users' reviews to generate explanations through structural approaches. Chang \etal \cite{chang2016crowd} propose an explanation model which learns tags from users, filling the generated tags into explanations written by a human. Similarly, Musto \etal \cite{musto2016explod} use a template text with predicted explainable properties to achieve good interpretations. In addition to constructing explanations in a modular way, Seo \etal \cite{seo2017interpretable} provide explanations by highlighting components in reviews.

Interpretations from those approaches are somewhat formulaic and do not consider personal styles of expression, since they are produced in a structured way. On the other hand, deep learning techniques have improved the machine generated text, especially recurrent neural networks (RNNs) \cite{sutskever2011generating}. RNNs are feed forward recursive neural network dealing with sequential information. Considering text generation as a sequence predicting problem, RNNs learn the patterns within a text, and predict the next token from the previous tokens. Machine generated text has made a significant breakthrough in simulating text writing, however the generation of high-quality synthetic reviews require the incorporation of more personalised information. Recently, many generated reviews incorporate individual information to enrich their personality and diversity \cite{lipton2015capturing}. Dong \etal \cite{dong2017learning}, show that generated reviews have great potential to make recommender systems more interpretable.

In this paper, we propose a review-oriented text generation framework that contains 3 popular generative models providing natural language oriented explanations for rating recommendation. More specifically, our generation models are based on two popular generative models:  character-level and word-level. Generative models usually learn attribute information from \textit{User ID}, \textit{Item ID}, and \textit{ratings}. We also note that not all human written reviews are consistent with their corresponding interactions, where the reviews have different characters, some focusing more on useful suggestions and others on personal opinions or experiences. It is common for review platforms (such as Amazon) to allow users to vote on the perceived helpfulness of the reviews, and this concept captures an amalgamated user view of some of these review attributes such as focus, conciseness and an accurate summary of the general opinion or sentiment about the item in question.
These judgments about the helpfulness of the reviews have a strong influence on users' purchasing decisions \cite{Danescu-Niculescu-Mizil:2009:WWW}.
Taking this observation into account, we learn from the helpfulness scores to achieve a stable consistency between generated reviews and recommendations. Our models are all based on RNNs architecture to learn alignments between reviews and attribute information. In order to evaluate our models, we create a dataset from Amazon books reviews \cite{mcauley2015image}. We first evaluate the generated text with natural language process (NLP) methods. These experiments show that each of our generation models outperform baseline methods, which demonstrates that the generated models have learned the item attributes well and can provide readable synthetic reviews to the user.

\begin{figure}
\includegraphics[width=0.5\textwidth]{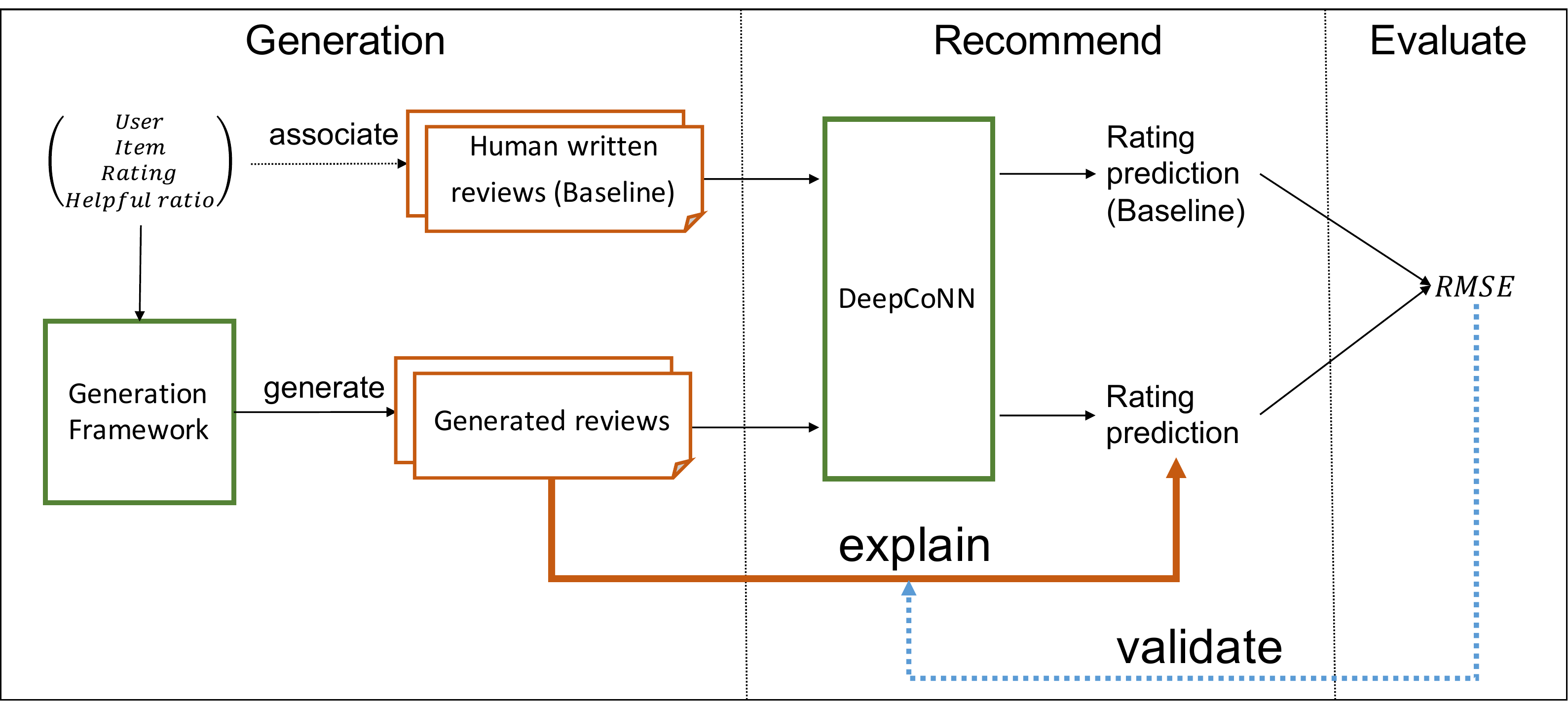}
\caption{Overview of the experimental setup to validate recommendation performance of generated reviews.}
\label{fig:rs_procedure}
\end{figure}

We then apply DeepCoNN \cite{zheng2017joint} method to measure and validate the recommendation performance of our generated reviews, as shown in Fig. \ref{fig:rs_procedure}. DeepCoNN is a review-oriented recommender system using reviews as inputs and ratings as output to recommend an item. The evaluation on the recommendation task is based on root mean square error (RMSE). Recommendation experiments demonstrate clearly that using synthetic reviews achieve better recommendation performance when compared to human written reviews. 
We investigate this process more closely by evaluating the readability of the synthetic reviews, and studying the correlations between the sentiments in the synthetic reviews and the ground truth ratings.

The contributions in this paper are as follows:

\begin{itemize}
\item We introduce the task of generative natural language explainable recommendation.
\item Our experiments illustrate machine generated reviews achieve better recommendation performance than human written reviews, which is an interesting approach to improve recommendation performance.
\item We provide the first machine-generated natural language explanations framework for rating recommendation.
\end{itemize}

\section{Related Work}
Recently, neural networks have been successfully applied to recommender systems, for example, \cite{donkers2017sequential, DBLP:conf/acml/KoMG16} who reported good performance applying recurrent neural networks (RNNs) for recommendation task. However, those state-of-art techniques suffers from the same problem as other recommender systems, which is a lack of interpretability. On the other hand, neural networks have recently shown improvements in the natural language processing (NLP) domain, such as text classification \cite{hughes2017medical}, sentiment analysis \cite{wang2018sentiment}, and text generation \cite{DBLP:conf/iui/CostaODL18, sutskever2011generating}. 

Almahiri \etal \cite{Almahairi:2015:LDR:2792838.2800192} developed one of the first recommender system models combined with NLP techniques, where they use review text as side information to improve the recommendation performance based on a RNN. Additionally, \cite{bansal2016ask}, \cite{zheng2017joint}, and \cite{chen2018neural} directly utilize reviews as inputs showing remarkable recommendation achievements by using convolutional neural networks (CNNs). Furthermore, \cite{seo2017interpretable} applied CNN to introduce an explainable recommender system on review level. 

These works validate the utility of user-generated reviews to enrich the performance of recommender systems. However, it is commonly observed that users' reviews contain not only direct experiences of the item, but often some irrelevant information which can mislead other users' and cause confusion or add noise about commonly understood features of the item. Generally, we expect this noise or misdirection to be reflected in the helpful votes by other users. The goal of our work is to generate natural language explanations which minimise the confusion and highlight the relevant users' sentiments to improve the recommendation task and show that synthetic reviews can be more useful as summarised explanations of the item features.

Sutskever \etal \cite{sutskever2011generating} is the first project applying large RNNs to generate text. The work of Karpathy \cite{karpathy2015visualizing} provides the first insights into context language generation model by using RNNs. Regarding review generation, according to Lipton \etal \cite{lipton2015capturing}, learning the rules for generating reviews can be accomplished by representing the input as characters or words. The word-level models invariably suffer from computational costs of an unfixed vocabulary list. Character-level models for review generation based on RNNs use concatenation methodology to learn auxiliary information. To reduce the computational complexity in word-level model as mentioned in \cite{lipton2015capturing}, Tang \etal \cite{tang2016context} introduce a context-aware model that uses only context information once rather than replicating them many times, and successfully generates word-level reviews. Furthermore, Dong \etal \cite{dong2017learning} propose an advanced version of the word-level model in \cite{tang2016context} by using the attention mechanism. 

Although these models show good improvements on the review generation task, their generated reviews use only the character or word level. It is not clear whether these models will demonstrate analogous text performance on both character and word level. Moreover, the manner of text generation will directly affect the text representations, and this leads to further divergence in the  explainability of the generated text. 
Thus, models in our framework generate explanations at both character and word level. Furthermore, as aforementioned, to eliminate misunderstanding of reviews, all our models are boosted by \textit{helpful ratio} to obtain useful information during the generation of explainable recommendation.

\section{Interpretation Models}

\begin{figure*}
    \centering
    \begin{subfigure}[t]{0.25\textwidth}
        \centering
        \includegraphics[width=\linewidth, height=3cm]{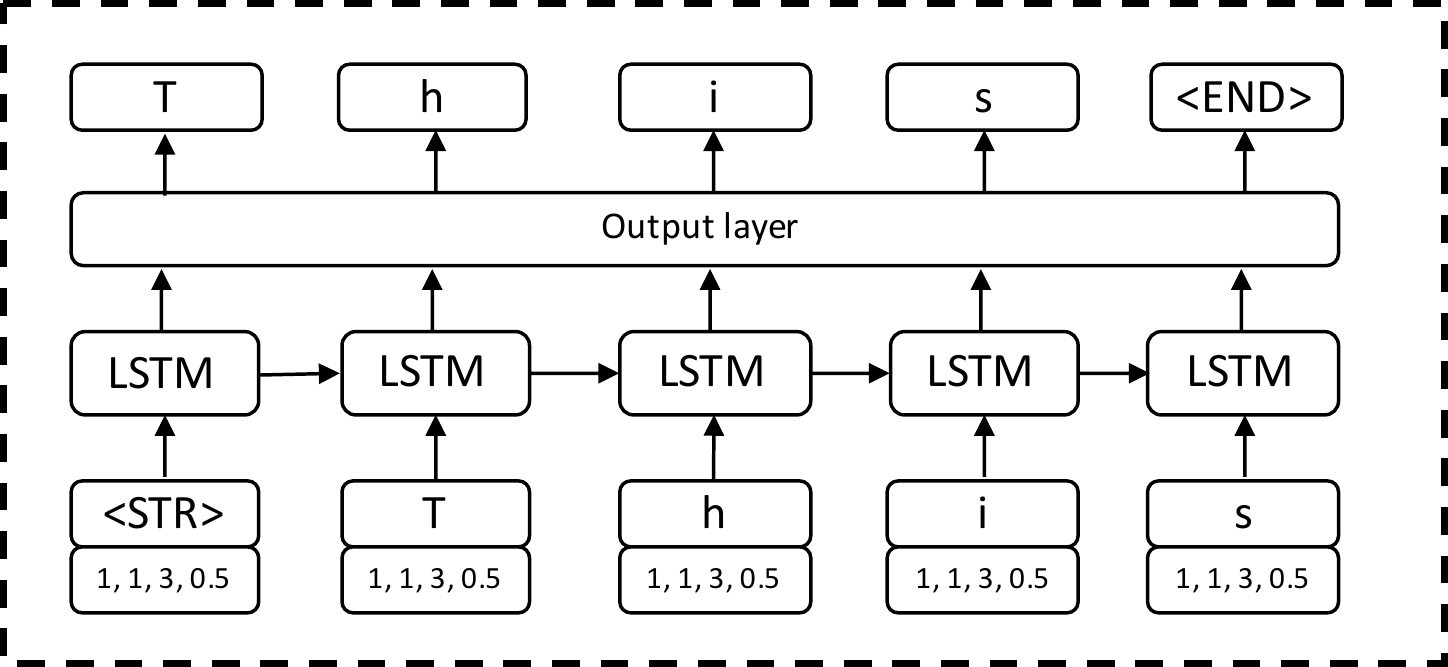}
        \caption{GCN generation network}
    \end{subfigure}%
    ~ 
    \begin{subfigure}[t]{0.37\textwidth}
        \centering
        \includegraphics[width=\linewidth, height=3cm]{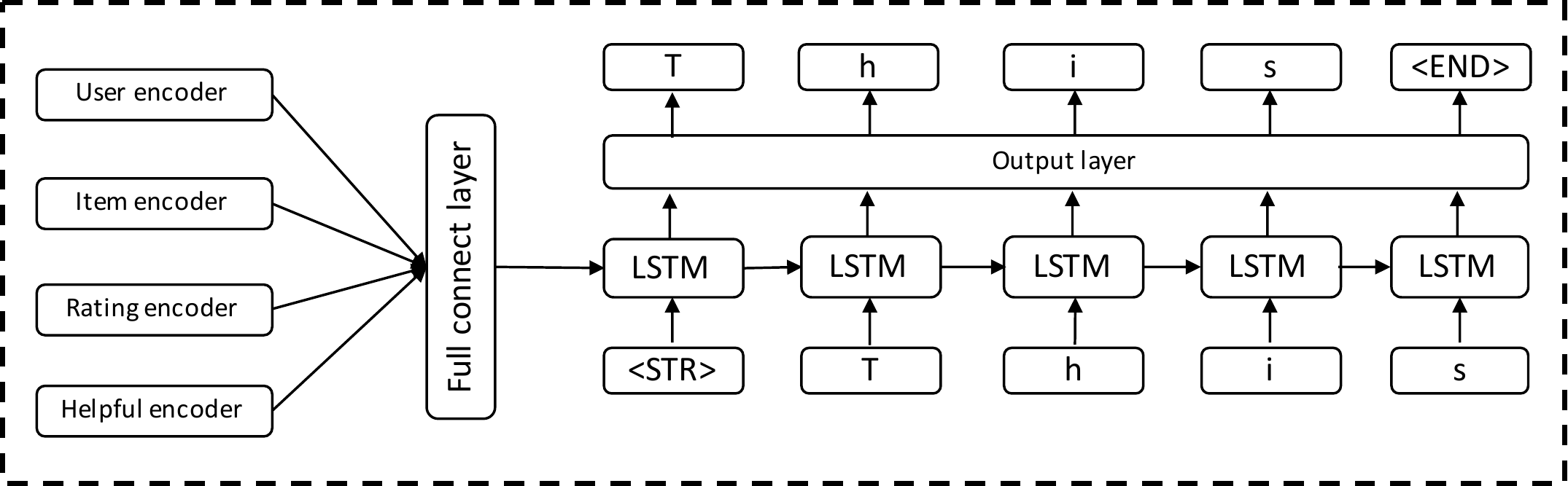}
        \caption{Context generation network}
    \end{subfigure}
    ~ 
    \begin{subfigure}[t]{0.38\textwidth}
        \centering
        \includegraphics[width=\linewidth, height=3cm]{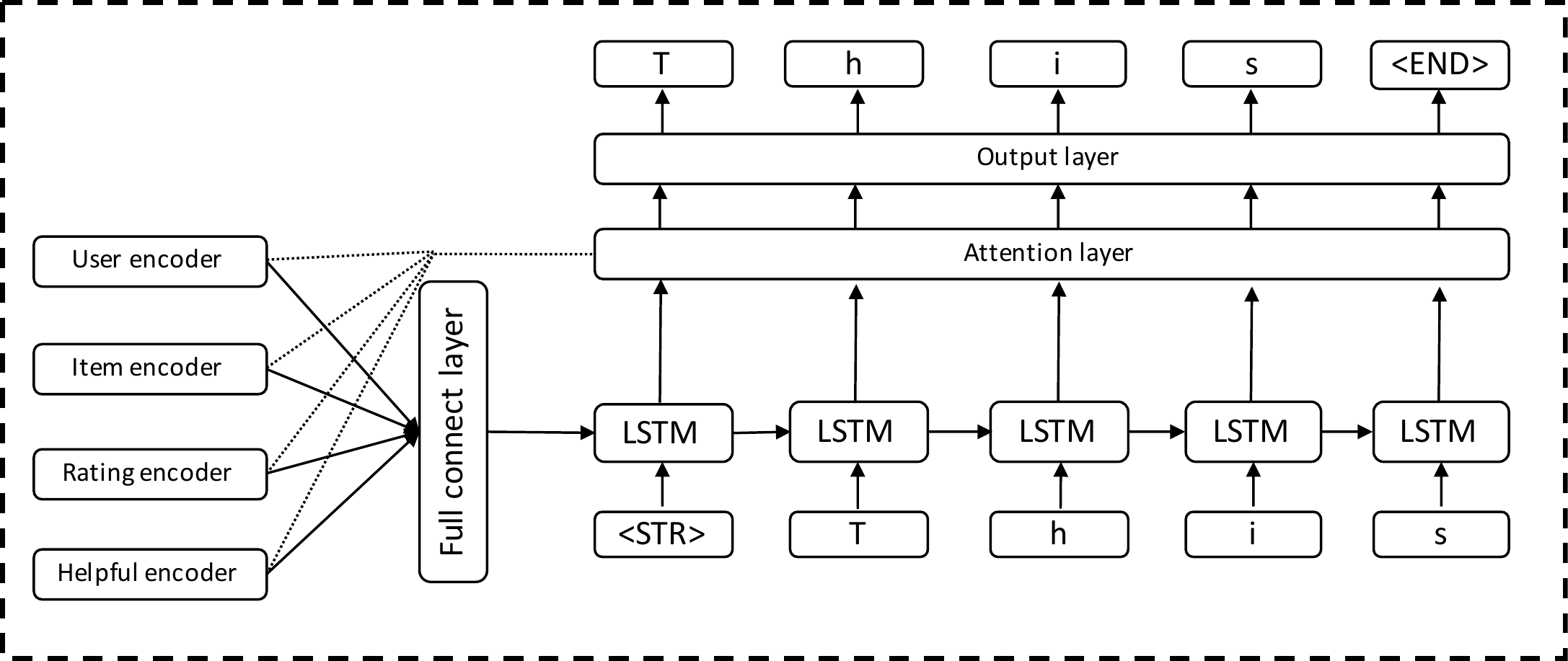}
        \caption{Attention generation network}
    \end{subfigure}
    \caption{(a): using concatenation approach on both sequence and attributes information; (b): multi-layer neural networks as encoder to initialise decoder network; (c): similar to (b), but stacking an attention mechanism into decoder network and output layer.}
    \label{fig:framework}
\end{figure*}

\subsection{Problem Statement}
The research problem investigated in this paper is defined as followed: How can we generate natural language explanations that meet the demand of rating recommendations? To answer this question our model not only learns regulation of text writing, but also which attributes suit the demand of recommender systems. We observe that the rating is associated with user-item pairs, while the helpfulness votes of a review reflect whether the review text has some utility to the other users. Therefore, we apply \textit{user ID}, \textit{item ID}, \textit{rating}, and \textit{helpful ratio} as the main attributes to learn in this task. \textit{helpful ratio} measures whether the review helps the user decide on the recommended item, being a ratio of number of helpful votes by total votes. As aforementioned, the text generation task is classified into two folds: the character and word level. We propose these two methods for each of our generative models to investigate the difference between character and word level on both text generation and recommender system tasks. While our generative models are different in the types of inputs and the underlying models, they aim to solve the same problem regarding the maximisation of the likelihood of the generated reviews.

We denote the input attributes vector as $a_i = (a_i, \dotso, a_{|a|})$, where $a$ is the attributes group consisting of related attributes, such as {user ID}, {item ID} {rating}, and {helpful ratio}. We aim to generate an explanation $e = (y_1, \dotso, y_{l})$,  where $l$ represents the sequence length. We generate an explanation by maximising the conditional probability $p(e|a)$. Finally, we formulate the generative steps as follows, where $y_{t} = (y_1, \dotso, y_{t-1})$.

\begin{equation}
\begin{aligned}
p(e|a) = \prod^{l}_{t=1} p(y_{t} | y_{<t}, a)
\end{aligned}
\label{fig:prob_state}
\end{equation}

\subsection{Models Framework}

Generative networks have been introduced to address the text generation problem, such as generative concatenative networks\cite{lipton2015capturing}, context-aware generation model\cite{tang2016context, dong2017learning}, and attention enhanced generating model\cite{dong2017learning}. However, as mentioned before, the generated text from these models is only focused on a single level, i.e. character or word. Meanwhile, they have not been verified on the recommendation task and do not take into account feedback from reviewers. Hence, we propose extended models from \cite{lipton2015capturing}, \cite{tang2016context}, and \cite{dong2017learning}, where each model generates text from both character and word level. All of our models take \textit{user ID}, \textit{item ID}, \textit{rating}, and \textit{helpful ratio} as inputs.

Fig \ref{fig:framework} demonstrates the three based models, and we refer them as the GCN model (generative contcatenative model), context model, and attention model. They assume identical techniques in the decoder section but are different in other parts. The GCN model learns the correlations of attributes and text by concatenating them as inputs. On the other hand, both context and attention models learn the alignment between attributes and the text by initializing decoder weights from the encoded attributes. During the generating step, both GCN and context models apply a single output layer to infer text, while the attention model stacks the output layer on an attention mechanism. The output vectors are interpreted as holding the $log$ probability of the next character in the sequence and the objective is to minimize the average cross-entropy loss over all targets. In the following sections, we introduce details of our models through a generating step.

\subsection{Encoder}

The three models we use assume different techniques in their processing methods, except in the decoding step. We generate a dictionary for all characters (or words) in the corpus to record their positions for character (or word) level, which will be used during the encoding process in the training step and for decoding in the generating step. Generally, encoder modules in our models can be categorised into two sections: encoding attributes and encoding text. During the text encoding process, the three base generative models encode characters and words by one-hot vectors and fixed length embeddings separately. On the other hand, for attributes encoding, context and attention model share the same methodology, while the GCN model uses a different approach.

The GCN model represents \textit{user ID} and \textit{item ID} by fixed length vectors similar to \cite{lipton2015capturing}, while one-hot vectors with maximum rating length and single continuous values stand for \textit{rating} and \textit{helpful ratio}, respectively. Thereafter, the encoded attributes $x^{a}$ are duplicated by text length, and concatenated with the encoded text $x^{r}$ to produce encoder output $X'_t$ for the GCN model, as shown in Eq. \ref{eq:concat}. Here, $t$ stands for time step $t$ in the text, $[:]$ denotes concatenation manipulation. $X'_t$ is then passed to decoder module as the input.

\begin{equation}
\begin{aligned}
X'_t = [x^{r}_{t}: x^{a}_{t}]
\label{eq:concat}
\end{aligned}
\end{equation}
In context and attention models, we apply multi-layer perceptron with one hidden layer to encode attributes into embeddings with a fixed dimension as seen in \cite{dong2017learning}. To do so, when receiving the one-hot representation of attribute $e(a_i)$, where $i \in (1, \dotso, |a|)$, we provide the encoded attribute $x^{a}_{i}$ in:

\begin{equation}
\begin{aligned}
x^{a}_{i} = W^{a}_{i} e(a_i),
\label{eq:gcn_encoder}
\end{aligned}
\end{equation}

where $W^{a}_{i} \in \mathbb{R}^{m \times |a_i|}$ is a weighting matrix, $m$ is the encoding dimension for all attributes, $|a_i|$ is length of attribute $i$. Unlike GCN model, we use encoded attributes to initialise decoder weights instead of using them as a part of decoder inputs. Thus, we concatenate encoded attributes and pass them into a fully connected layer supplying an output to have the same shape with decoder weights, as shows in:

\begin{equation}
\begin{aligned}
A = \tanh (H [x^{a}_{i}, \dotso, x^{a}_{|a|}] + b_{a}),
\label{eq:context_attention_encoder}
\end{aligned}
\end{equation}

where, the full connected layer uses $\tanh$ activation function, $H$ is a parameter matrix, and $b^a$ denotes the bias.

\subsection{Decoder}

Recurrent neural networks (RNNs) are applied in the decoder module as used in other text generative models \cite{lipton2015capturing, tang2016context, dong2017learning}. RNNs are feed-forward networks with dynamic temporal behaviour aiming to process and learn sequential data. Regarding the generation task, RNNs summarise context information into hidden variables and then provide probability distributions for each time step. In a conventional RNN, given an input vector $X_t$ during time step $t$ and the cell state of previous time step $t-1$, it performs a $\tanh$ transmission to get a hidden state of time $t$. We make a prediction of time $t$ by passing the hidden state to an output layer activated by a non-linear $softmax$ function, as shown in Eq. \ref{eq:rnn}.

\begin{equation}
\begin{aligned}
& h_t = \tanh(W_{x} X_t + W_{h} h_{t-1}) \\
& p(y_{t} | y_{<t}, a) = \mathrm{softmax}(W h_t + b),
\end{aligned}
\label{eq:rnn}
\end{equation}

This mechanism helps conventional RNN to learn the sequential reliance of the input data. However, it always gets lost and forgets previous information over time- this is the well-known gradient vanishing problem, a common issue with conventional RNN's. To solve this issue, \cite{hochreiter} introduced long short-term memory (LSTM) cells, consisting of a set of gates: forget $f$, input $i$, and output $o$. They define respectively which section to be discarded, which information needs to be remembered, and which knowledge should pass to the rest of network. Similarly to conventional RNNs, forward calculation of an LSTM unit involves inputs $x_t$, cell state $C_{t-1}$ from previous unit, and previous unit output $H_t$. Formal calculation steps are defined as:

\begin{equation}
\begin{aligned}
& \hat{C}_t = \tanh (W^c_x x_t + W^c_h H_{t-1} + b_c) \\
& f_t = \sigma(W^f_x x_t + W^f_h H_{t-1} + b_f) \\
& i_t = \sigma(W^i_x x_t + W^i_h H_{t-1} + b_i) \\
& C_t = f_{t} \odot C_{t-1} + i_{t} \odot C'_{t} \\
& o_{t} = \sigma(W^o_x x_t + W^o_h H_{t-1} + b_o) \\
& H_t = o_{t} \odot \tanh(C_t) \\
\end{aligned}
\label{eq:lstm}
\end{equation}

where $W$ and $b$ stand for weights and bias respectively, $\hat{C}$ is candidate cell state, and $\odot$ denotes an element wise product.

\subsection{Text Generation}

Text generation is described as a sequence label classification problem. The GCN and context model apply a single output layer to generate text, while the attention models employ an attention layer before the output layer. We propose the same attention mechanism as \cite{dong2017learning}, but using more attributes to enhance the alignment between context and attributes values.

To do so, we feed the decoder output $H_t$ to an output layer activated by a $Softmax$ function computing conditional probabilities $p(y_{t} | y_{<t}, a)$ for all characters (or words) at time $t$ as \cite{lipton2015capturing}. Then, the generation module maximises the conditional probabilities $p(y_{t} | y_{<t}, a)$ by a greedy search function to infer the index $Y_t$ of the generated character (or word). Thereafter, we infer a character (or word) by looking up $Y_t$ in the dictionary created previously. This procedure is applied recursively and a group of characters (or words) is generated until we find the pre-defined $end$ symbol. The calculation steps of this procedure are presented as:

\begin{equation}
\begin{aligned}
& p(y_{t} | y_{<t}, a) = \mathrm{softmax}(W H_t + b) \\
& Y_t = \argmax p(y_{t} | y_{<t}, a)
\end{aligned}
\label{eq:seq_decoder}
\end{equation}

\section{Experimental Evaluation}
\label{sec:experimental_evalation}
In this section, we first present details of our dataset, the experimental settings and the baseline methods and evaluation metrics. Then we demonstrate cross-domain experiments from natural language process (NLP) to the recommender systems domain. We first evaluate the correspondence between the generated text and that of human written text in Section \ref{sec:coherence}. Section \ref{sec:readability} analyses the comprehensibility of our generated reviews by comparing them with baseline reviews through readability scores. In Section \ref{sec:perfrecsys}, we compare and explain the performance of the generated reviews and baseline reviews on a fine-tuned recommendation system. We measure to what extent the generated reviews make a contribution to the recommendation task, and why generated reviews provide such a boost in performance. In order to investigate the reason why generated reviews present better recommendation performance than human written reviews, we have studied the correlation relationship between the reviews' sentiments and ground truth ratings in Section \ref{sec:sentiment}. In this section, we have discussed the feasibility and rationality of using generated reviews to explain a recommended rating. We also investigate how synthetic reviews can carry more consistent information appropriate to the demands of a recommender system than human written reviews.

\begin{figure*}
    \centering
    \begin{subfigure}[t]{0.5\textwidth}
        \centering
        \includegraphics[width=\linewidth, height=9cm]{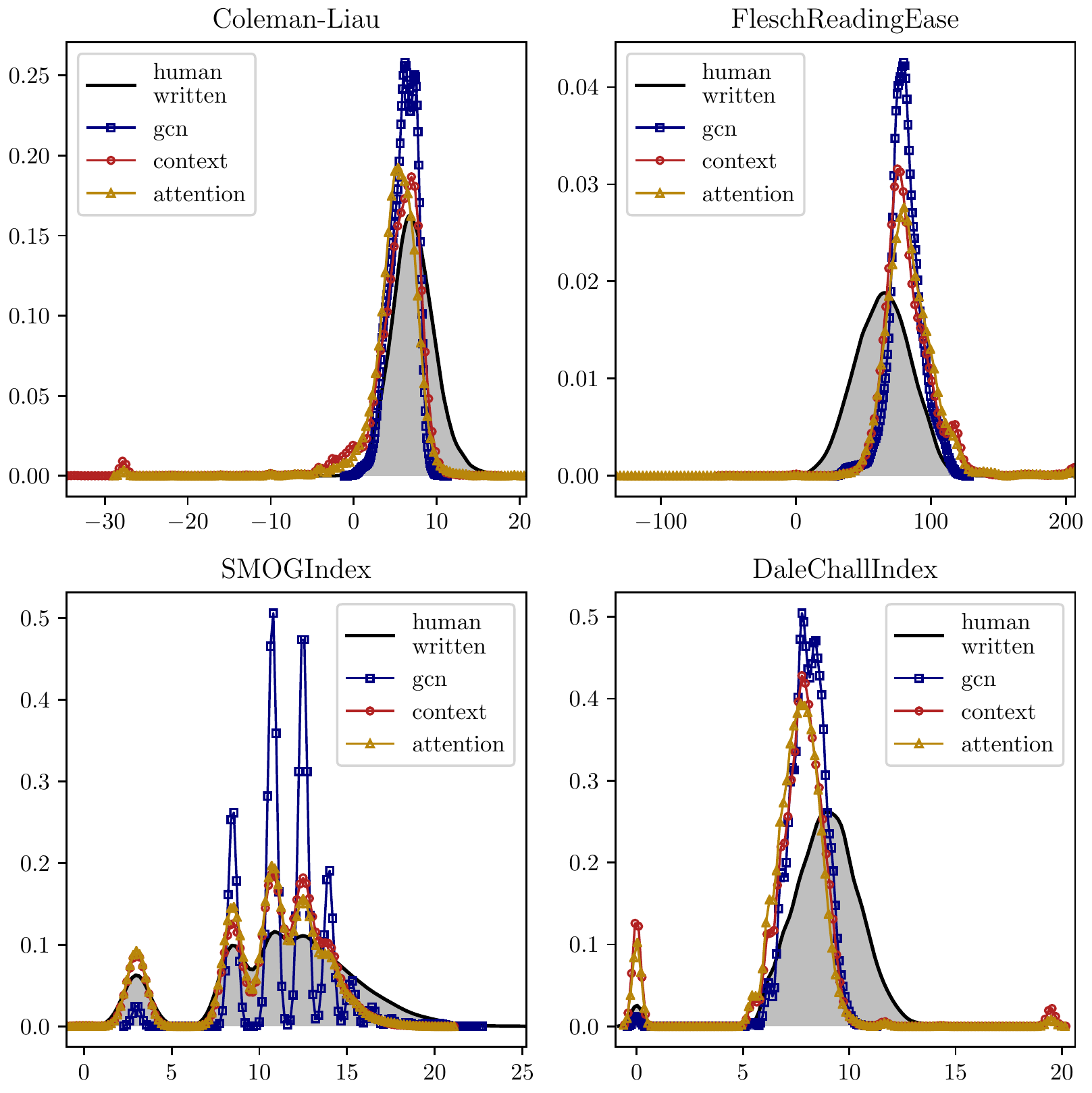}
        \caption{Character-level}
        \label{fig:reada}
    \end{subfigure}%
    ~ 
    \begin{subfigure}[t]{0.5\textwidth}
        \centering
        \includegraphics[width=\linewidth, height=9cm]{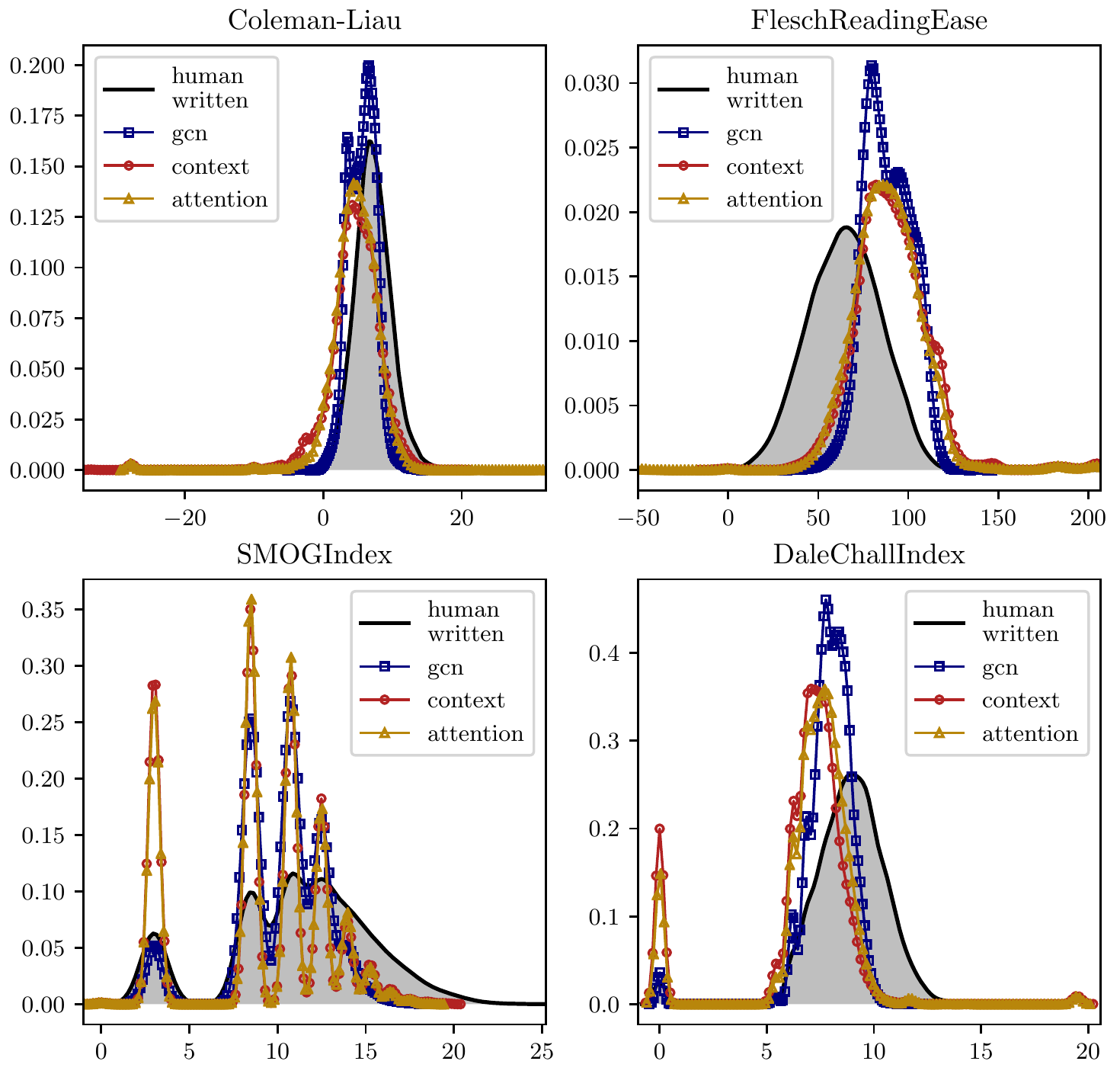}
        \caption{Word-level}
        \label{fig:readb}
    \end{subfigure}
    \caption{Readability Evaluation}
    \label{fig:read}
\end{figure*}

\subsection{Experimental Setup}

\subsubsection{Data Preparation}
We build a review dataset with helpfulness ratio from the Amazon product data \cite{mcauley2015image}. This dataset contains unique reviews spanning the period from May 1996 to July 2014. We focus on data from the books domain in this task. Primary challenges of this task are whether generated reviews can improve rating predictions. To answer this questions, the dataset should consist of user-item interactions. According to Dong \etal \cite{dong2017learning}, is is observed that long reviews in the books dataset tend to focus on summarising the book itself and are not rich in user experiences which are of more interest for our task. Therefore, reviews whose lengths are greater than 70 words are filtered out. We keep the reviews which have at least one helpful vote, as reviews without a helpful vote would not reflect an influence on the users' decisions. Thereafter, we calculate the helpful ratio by dividing number of helpful votes by total votes. We keep books and users which both appear no less than 5 times to learn diverse expressions. The dataset contains 284,545 reviews paired with four attributes, i.e. {user ID}, {item ID}, {rating}, and {helpful ratio}. Additionally, we have 30,698 users and 21822 items. Ratings are discrete values from 1 to 5, while helpful ratios are continuous values from 0 to 1. The average review length is around 40 words and 216 characters. The total number of words in this dataset is about 111,000. We then randomly split this dataset into a \textit{TRAIN} set, a \textit{VALIDATE} set, and a \textit{TEST} set in the proportion of 70\%, 10\%, and 20\% respectively.

\subsubsection{Experiment Settings}

Our models learn in two way to generate reviews: with a word-level and character-level approach. We use a regular expression to tokenise the reviews, filtering out tokens which occur less than 16 times. Words are represented by embeddings whose dimensions are set to 512 and a one-hot vector with a length of 100 for characters. Regarding the attributes values, i.e. users, items, ratings, and helpful ratios, the representation of those are different depending on the model. The GCN model represents users and items by vectors with a dimension of 64, while a one-hot vector and a single value stand for the ratings and helpful ratio respectively. Context and attention model use vectors with a dimension of 64 for those attributes. The models are built on the same decoder architecture: that of a stacked two-layer RNN with LSTM units. We initialise all parameters of our models from a uniform distribution in range of [-0.08, 0.08] as suggested by \cite{karpathy2015visualizing}. We employ RMSEProp \cite{bengio2015rmsprop} optimisation tuning models with initial learning rate 0.002 and decay 0.95. Our networks are unrolled for 100 epochs. We decrease the learning rate in every epoch after 10 epochs by multiplying with a factor of 0.95 to avoid over-fitting. Then we stack a dropout layer on each hidden layer with keep probability of 0.8. Then, we clip gradients in a range of [-5, 5] to avert the gradient exploding problem \cite{pascanu2013difficulty}. We select the best model based on the results of the \textit{VALIDATE} set. We compose a review by feeding a character/word in time step $i-1$ to predict a character/word in time step $t$ during the generation step. We use a greedy search methodology to select a character/word during a time step $i$. We use the DeepCoNN model \cite{zheng2017joint} to analyse our generated reviews on recommendation task. We split the user-item pairs in \textit{TEST} set into another \textit{train}, \textit{validate}, and \textit{test}. The \textit{train} set is used for training the DeepCoNN model. The \textit{validate} set aims to select best hyper-parameters. The \textit{test} set measures the performance of the \textit{TEST} set on the recommendation side. We then replace reviews in the \textit{TEST} set by our generated reviews, and use the user-item pair in \textit{test} to evaluate the behaviour of generated reviews for recommendation.

\subsubsection{Baseline Methods}
We compared our results with different baselines from the NLP field to the recommendation system in a progressive way. The baseline methods of the NLP domain are:

\begin{enumerate}
\item \textbf{Rand} The predicted reviews are arbitrarily sampled from \textit{TRAIN} as in \cite{dong2017learning}. This method provides the minimum NLP standard for this task.
\item \textbf{User-NN} This method uses the nearest neighbour method that randomly selects reviews from \textit{TRAIN}, where users and ratings of these reviews are the same as users and ratings from \textit{TEST} set.
\item \textbf{Item-NN} The same strategy as \textbf{User-NN} but retrieves reviews by the same item and same rating.
\end{enumerate}

Thereafter, we evaluate the performance of the recommender system with our generated reviews and compare with the following baseline:

\begin{enumerate}
\item \textbf{Test-pair} This approach consists of \textit{test} user-item pair and corresponding human composed reviews in \textit{TEST}.
\end{enumerate}

\subsubsection{Evaluation Metrics}

Our task is composed of cross-domain problems. Therefore, we use different metrics for distinct problems. We introduce them through our validation steps:

\begin{enumerate}
\item \textbf{BLEU} Bilingual evaluation understudy (BLEU) score \cite{papineni2002bleu} is an algorithm which measure the correspondence between the synthetic outputs and that composed by a human. It has  been  shown
to correlate well with human judgment on many text generation tasks \cite{dong2017learning}.
\item \textbf{Readability} To evaluate whether a generated review is readable or not we use 4 readability metrics to inspect the coherence of readability distribution between reviews in \textit{TEST} and generated reviews: Coleman Liau index \cite{Pera2}, Flesch reading ease \cite{Pera1}, simple measure of gobbledygook (SMOG) \cite{smog}, and Dale Chall \cite{dale1948formula}.
\item \textbf{RMSE} We calculate Root Mean Square Error (RMSE) to evaluate the models' performance on recommendation task. It is a popular evaluation method in the rating regression problem, especially rating prediction \cite{zheng2017joint}. Generally, a lower RMSE means a better performance. RMSE is defined as:

\begin{equation}
\begin{aligned}
RMSE = \sqrt{\frac{1}{N} \sum{u, i} (\hat{R}_{u, i} - R_{u, i})^2}
\end{aligned}
\label{eq:rmse}
\end{equation}

For a particular user-item pair $(u, i)$, $\hat{R}_{u, i}$ represents their predicted rating, while $R_{u, i}$ stands for their real rating. $N$ is the total number of user-item pairs.

\item \textbf{Rating discrepancy} The effectiveness of an explanation in the recommender system can be measured by the discrepancy between ground truth rating and rating on the basis of synthetic generated text explanations as introduced by \cite{tintarev2012evaluating}. The definition of the rating discrepancy is:

\begin{equation}
\begin{aligned}
Delta_{i} = \hat{\mathcal{N}}_{i} - \mathcal{N}_{i}
\end{aligned}
\label{eq:delta}
\end{equation}

where, $\hat{\mathcal{N}}_{i}$ stands for the predicted value of element $i$, while $\mathcal{N}_{i}$ denotes the value of element $i$ in the actual distribution.

\item \textbf{Pearson correlation} Pearson correlation is a well known method to measure linear correlations between two examples. The range of this metric is between -1 and 1, where -1 denotes negative linear correlation, 0 indicates absence relationship, and 1 represents positive linear correlation. We can formally define the Pearson correlation shows as:

\begin{equation}
\begin{aligned}
Pearson = \frac{{}\sum_{i=1}^{n} (x_i - \overline{x})(y_i - \overline{y})}
{\sqrt{\sum_{i=1}^{n} (x_i - \overline{x})^2(y_i - \overline{y})^2}}
\end{aligned}
\label{eq:pearson}
\end{equation}

where $x$ and $y$ are the two examples for comparison.

\end{enumerate}

\subsection{Coherence with Attributes} \label{sec:coherence}
The initial experiments using BLEU-4 score metric aim to validate the analogy between generated texts and human written reviews. According to \cite{dong2017learning}, BLEU-4 score is used to measure the precision up to 4 grams by comparing the generated texts with human written reviews, and penalises length using a brevity penalty term. Using the BLEU-4 score we can infer the correlation between the generated text and the original review. Hence, we evaluate the generated texts from our methods and the three baselines methods previously defined to compare the attributes coherency with corresponded \textit{TEST} reviews. Note, the difference between text and reviews is that reviews provide latent pieces of information that relate to attributes, while the general text does not.

\begin{figure}
\includegraphics[width=0.5\textwidth]{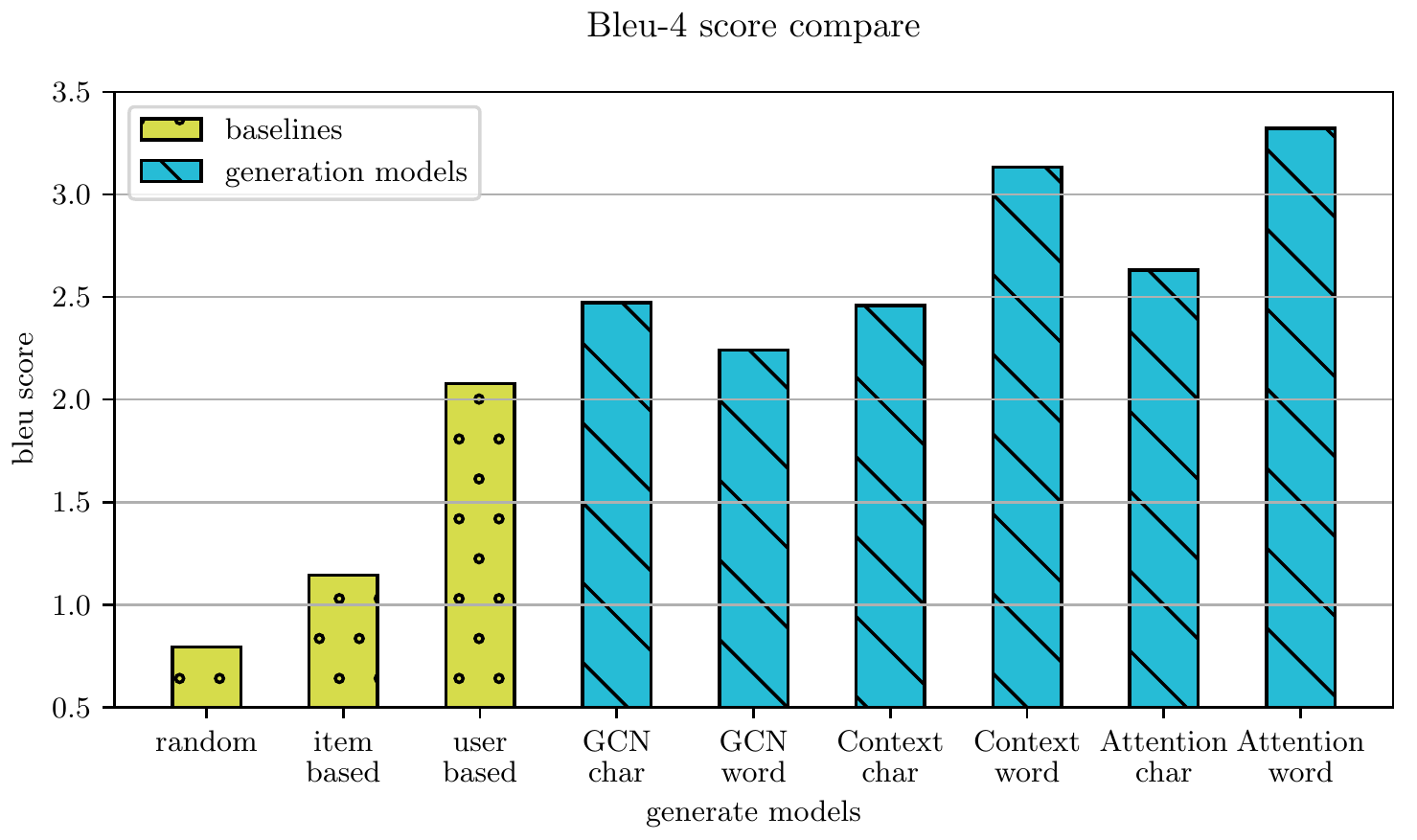}
\caption{BLEU evaluation on \textit{TEST} set}
\label{fig:bleu}
\end{figure}

According to Fig. \ref{fig:bleu} we observe that the random method presents the lowest score, due to randomly selecting reviews from training set as the generated reviews do not correlate well with their attributes. The item-based nearest neighbourhood method achieves a higher score than the random algorithm, however it does not perform well in comparison to other methods, since it uses reviews and ratings from the same item. The user-based nearest neighbourhood algorithm performs better than the previous methods because it applies reviews and ratings given by a user, making a personalised prediction. Analysing the results from our generative models we observe they outperform the naive baselines, due to their ability to combine information from \textit{User ID}, \textit{Item ID}, \textit{ratings}, and \textit{helpful ratio} to generate the texts. This is a compelling justification to identify these generated texts as synthetic reviews. 

\subsection{Readability Analysis}\label{sec:readability}
In order to measure the readability and comprehensibility of generated reviews, we apply the 4 readability metrics on generated corpus and corresponding human written reviews in \textit{TEST}, and further plot the distribution of those readability values as shown in Fig. \ref{fig:read}.

Fig. \ref{fig:read} shows the readability distributions from 6 generated texts and \textit{TEST} reviews, whereas the distribution with shadow is from \textit{TEST} reviews and unshaded distribution comes from machine-generated texts. The title of each graph represents the used readability metrics. The Fig. \ref{fig:reada} shows the results comparing character based models and \textit{TEST} reviews, while unshaded distributions in the Fig. \ref{fig:readb} come from word-based models. According to these results, the generative models all work similarly across all the readability metrics. In the Coleman Liau index metric, both character-level and word-level models present the nearly same distribution with the \textit{TEST} review. However, in the SMOG index result \textit{TEST} reviews perform an inimitable distribution, both character-level and word-level learned in a similar manner and performed comparably. In other readability metrics we have evaluated, the distributions of generated texts do not match the \textit{TEST} reviews distribution all that well, but they are close and share the same scope. These experimental results show strong consistency arguments for the comprehensibility of the synthetic reviews.

\subsection{Performance on Recommender System}\label{sec:perfrecsys}
\begin{table}[b]
\begin{tabular}{@{}clc@{}}
\toprule
\textbf{Language Models} &  & \textbf{Root mean square error} \\ \midrule
GCN char                 &  & 1.15765                         \\
\textbf{GCN word}        &  & \textbf{1.15583}                \\
Context char             &  & 1.16548                         \\
Context word             &  & 1.16928                         \\
Attention char           &  & 1.15842                         \\
Attention word           &  & 1.16404                         \\
\textbf{Test-pair}       &  & \textbf{1.22605}                \\ \bottomrule
\end{tabular}
\caption{RMSE results}
\label{tab:rmse}
\end{table}

Our goal is to identify whether generated reviews can be used for rating recommendation explanations. As Zheng \etal \cite{zheng2017joint} show, reviews provide additional information which could improve the rating prediction. Assuming the generated reviews on a rating recommender system can achieve a reasonable level of performance, then the latent information of generated reviews meets the requirements of the recommender system. Accordingly, if the latent information from generated reviews shows acceptable correlations with ratings, such reviews can be used to explain the corresponded ratings. 

To test these ideas, we first compare the generative models with the Test-pair baseline by applying related user and item reviews to the DeepCoNN recommender system to measure the performance of the generation models on the recommendation task. Considering the results in Table \ref{tab:rmse} we observe that the Test-pair baseline delivers similar performance as shown in \cite{zheng2017joint}, which verifies that the DeepCoNN recommender system has accurately modelled the latent representation of both users and items. However, the proposed generative models perform better than the Test-pair baseline. The GCN models reach the best performance among our generative models. The effectiveness of the attention mechanism is observed, since the attention models present lower error than the context models in both character and word level. The best RMSE score of the generation models leads to a \textbf{5.3\%} improvement over the Test-pair baseline. These results indicate that synthetic reviews present more abundant information that fits the requirement of the recommender system than human written reviews. 



\begin{figure}
\includegraphics[width=0.5\textwidth]{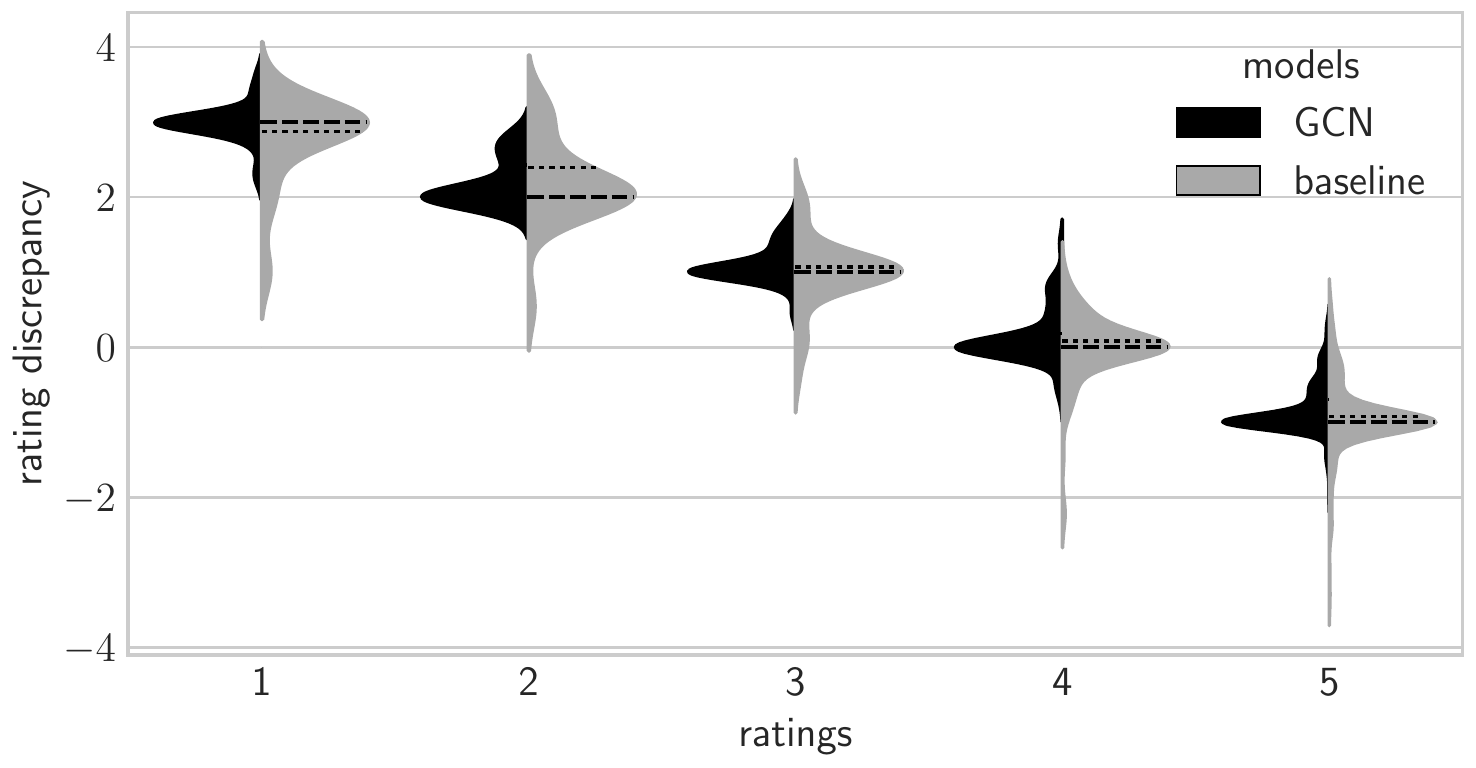}
\caption{Rating discrepancy comparison}
\label{fig:delta}
\end{figure}

The results in Table \ref{tab:rmse} are justified according to Eq. \ref{eq:delta}, which describes the rating discrepancy between the actual and predicted ratings for character-based GCN and the Test-pair baseline. In Fig. \ref{fig:delta} we analyse the rating discrepancy for each rating value from 1 to 5. The baseline and GCN models both make overestimates for reviews with lower ratings, and underestimate the reviews with highest rating (5). The average discrepancy is minimal for the most common rating value of 4. Although the GCN models have a similar distribution of predicted ratings to the baseline, the baseline clearly has more extreme outliers at both ends of the distribution. By filtering out the outliers across all rating values, the use of synthetic reviews improves on the state of the art rating prediction baseline model.

\subsection{Sentiment Consistency}\label{sec:sentiment}
Measuring the consistency between the rating prediction and explainable recommendation is not an easy task. Thus, we computed the polarity values of the generated reviews and the reviews from Test-pair baseline separately to understand whether the relationship between ratings and explanations are plausible. To do so, we used the Pearson correlation  to measure the extent of linear correlations between those sentiment scores and ground truth ratings. We observe in previous experiments that word-level models perform similarly to character-level models on DeepCoNN, therefore an evaluation on the character-level models is enough to analyse this consistency. 

The results of the Pearson correlations are shown in Tab \ref{tab:pearson}, where 100\% means perfect linear correlation. According to the results, the ground truth ratings have higher correlations with generated reviews than reviews in Test-pair. The GCN model shows strong linear correlations with ground truth ratings. Consequently, the results indicate a linear relationship between the polarity of generated reviews and ground truth ratings. The correlation values point to the reason why the generated reviews can reduce the influence of outliers. This validates the observation that synthetic reviews from generation models provide more consistent information for recommendations than human written text.
\begin{table}
\begin{tabular}{@{}cc@{}}
\toprule
\textbf{Models}        & \textbf{Pearson correlation (\%)} \\ \midrule
\textbf{GCN char}               & \textbf{79.50}                   \\
GCN char helpful       & 74.11                            \\
Context char           & 50.65                            \\
Context char helpful   & 63.42                            \\
Attention char         & 48.39                            \\
Attention char helpful & 58.01                            \\ 
\textbf{Test-pair}              & \textbf{44.69}                            \\ \bottomrule
\end{tabular}
\caption{Pearson correlation between polarity of reviews and related ratings (\textit{p} < 0.05)}
\label{tab:pearson}
\end{table}

\section{Conclusions}

In this paper, we propose a novel framework to provide rating recommendation explanations. We build 6 explanation generation models in this framework, and analyse their performance on both natural language generating and recommendation interpretation tasks. In natural language evaluation evaluations, readability metrics show strong comprehensibility of the generated text, while BLEU scores reveal that generation models have learned well the features of all attributes. Regarding the recommendation task, the RMSE results demonstrate \textbf{5.3\%} improvement over a state of the art recommendation baseline. This demonstrates that generated reviews provide more useful information than human writing reviews, which suits the particular demands of explainable recommender systems. Finally, we use the Pearson correlation to investigate the correspondence between review sentiments and ratings. Character-level generation models all outperform the recommendation baseline in this experiment, while GCN models show high linear correlation with ratings. According to our findings, generated reviews express more relevant experiential information about ratings. 
Additionally, word level models show similar performance to character-level model across all experiments.

This work suggests several interesting directions for explainable recommender systems. Interpretable recommender systems could learn more attributes, and provide better performance on both recommendation and explanation. Moreover, since natural language explanation models rely on models for generating text, more state of the art generation models, such as generative adversarial networks, could be employed to further improve these systems.

\bibliographystyle{ACM-Reference-Format}
\bibliography{reference}

\end{document}